\documentclass[twocolumn,twoside,slac_two,nofootinbib]{revtex4}

\usepackage{graphicx}
\usepackage{fancyhdr}
\usepackage{hyperref}
\hypersetup{
    colorlinks=true,
    urlcolor=magenta,
}
\begin{document}

\title{TRAGALDABAS. First results on cosmic ray studies and their relation with the solar activity, the Earth's magnetic field and the atmospheric properties\\
\vspace{5mm}
{\it The TRAGALDABAS Collaboration:}
}
%

%
\author{J. A. Garz\'on (juanantonio.garzon@usc.es)\\
 J. Collazo, J. Cuenca-Garc\'ia, D. Garc\'\i a Castro, J. Otero, M. Yermo}
\affiliation{LabCAF-IGFAE, Univ. Santiago de Compostela. 15782 Santiago de Compostela, Spain}
\author{J.J. Blanco}
\affiliation{Univ. Alcal\'a. 28801 Alcal\'a de Henares, Spain}
\author{T. Kurtukian}
\affiliation{CEN-Bordeaux, Univ. Bordeaux. F-33175 Gradignan, France}
\author{A. Morozova, M.A. Pais}
\affiliation{CITEUC-Univ. Coimbra. 3040-004 Coimbra, Portugal}
\author{A. Blanco, P. Fonte, L. Lopes}
\affiliation{LIP-Coimbra. 3004-516 Coimbra, Portugal}
\author{G. Kornakov}
\affiliation{TU Darmstadt. 64277 Darmstadt, Germany}
\author{H. \'Alvarez-Pol, P. Cabanelas, A. Pazos, M. Seco}
\affiliation{IGFAE, Univ. Santiago de Compostela. 15782 Santiago de Compostela, Spain}
\author{I. Ri\'adigos, V. P\'erez Mu\~nuzuri}
\affiliation{GFNL, Univ. Santiago de Compostela. 15782 Santiago de Compostela, Spain}
\author{A. G\'omez-Tato, J.C. Mouri\~no, P. Rey}
\affiliation{CESGA - Xunta de Galicia. 15706 Santiago de Compostela, Spain}
\author{J. Taboada}
\affiliation{Meteogalicia - Xunta de Galicia. 15707 Santiago de Compostela, Spain}

\begin{abstract}
Cosmic rays originating from extraterrestrial sources are permanently arriving at Earth's atmosphere, where they produce up to billions of secondary particles. The analysis of the secondary particles reaching to the surface of the Earth may provide a very valuable information about the Sun activity, changes in the geomagnetic field and the atmosphere, among others. In this article, we present the first preliminary results of the analysis of the cosmic rays measured with a high resolution tracking detector, TRAGALDABAS, located at the Univ. of Santiago de Compostela, in Spain.
\end{abstract}

\maketitle

\thispagestyle{fancy}


\section{Introduction}

\begin{figure*}[t]
\centering
\includegraphics[width=160mm]{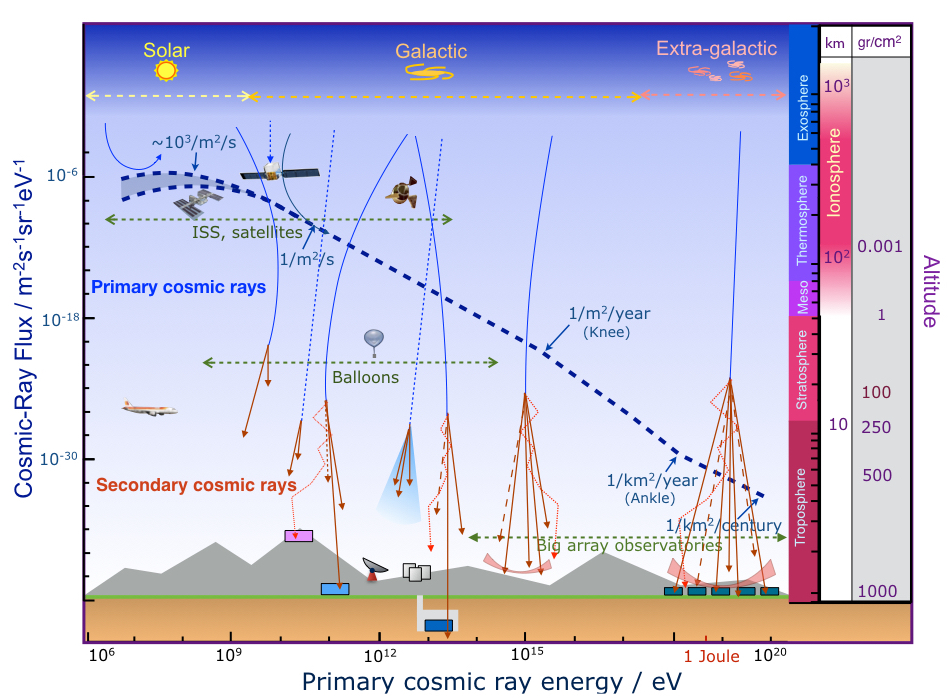}
\caption{
Primary cosmic ray spectrum as a function of the energy. The blue dashed line represents the cosmic rays flux using the scale units shown on the left vertical axis. Below energies $\sim 10^{14}$ eV primary cosmic rays are measured directly with detectors placed in satellites and balloons. At higher energies indirect measurements done by ground-based arrays of detectors are used. The right vertical axis shows the atmosphere layers, their heights and the corresponding air pressure.
}
\label{f_crspectrum}
\end{figure*}

More than one hundred years after the discovery of cosmic rays there are still many unknowns such as how they reach energies up to more than $10^{20}$ eV, what is their origin or their composition at the highest energies, among others. One of the difficulties inherent to the study of the high energy cosmic rays is that those of them having energies above $\sim 10^{15}$ eV are so scarce that their properties can only be estimated analyzing the arrival distribution of the extended air showers (EAS) of secondary particles using ground-based detectors.  Since, at the highest energies, EAS are spread-out over very wide surfaces, a significant sample of the arriving particles is typically obtained using huge arrays of low granularity counting devices, such as plastic-scintillators or water Cherenkov detectors with a modest time resolution. As a consequence, ironically, the highest energetic particles ever seen are measured with a few of the simplest and cheapest available detectors. Figure \ref{f_crspectrum} summarizes most of the features of the primary cosmic ray spectrum together with their possible sources and the common detection methods.

\begin{figure*}
\centering
\begin{minipage}{.49\textwidth}
  \centering
  \includegraphics[width=0.9 \linewidth]{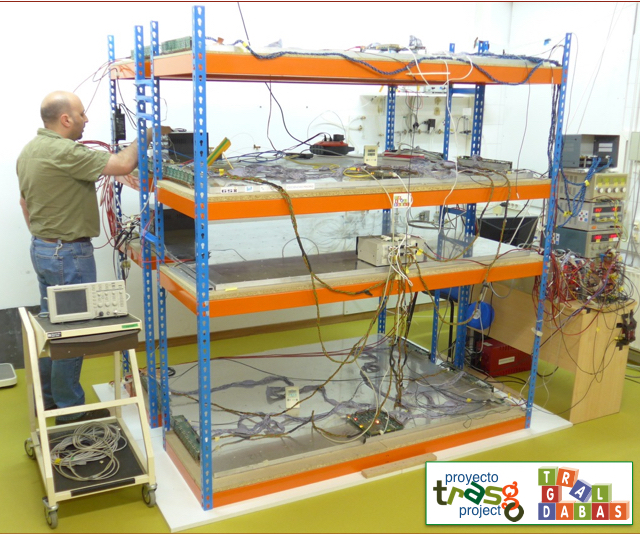}
  \caption{Picture of the TRAGALDABAS detector, at the Faculty of Physics of the Univ. of Santiago de Compostela, in Spain.}
  \label{f_tragaldabas}
\end{minipage}\hfill%
\begin{minipage}{.49\textwidth}
  \centering
  \includegraphics[width= 0.9 \linewidth]{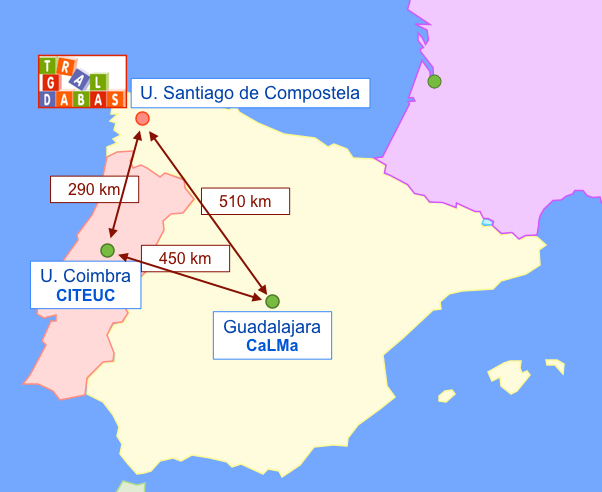}
  \caption{Location of the TRAGALDABAS detector together with the CITEUC, in Coimbra (Portugal), and CaLMa, in Guadalajara, in the Iberian Peninsula.}
  \label{f_mapa}
\end{minipage}\hfill%
\end{figure*}

In order to improve the understanding of the space-time properties of the front region of cosmic ray air showers, we have built and installed in the Faculty of Physics at the Univ. of Santiago de Compostela, in Spain, a high position and time resolution tracking detector of the TRASGO family\cite{b_trasgo}, called TRAGALDABAS\cite{b_tragaldabas}. Its present layout is shown in Figure \ref{f_tragaldabas}.
The detector aims to go a step forward  into the analysis of some of the effects related to the microstructure of the EAS, observed during the commissioning of the tRPC (timing Resistive Plate Chamber) detectors of the HADES ToF wall\cite{b_hadesmicrostructure} at the GSI laboratory. 
A very singular circumstance of our detector, is that it is relatively close to the Center for Earth and Space Research of the University of Coimbra, CITEUC\cite{b_citeuc}, and the CaLMa neutron monitor\cite{b_calma}. The location of both centres and their distances to TRAGALDABAS are shown in Figure \ref{f_mapa}. Several features of the detector have been already published in\cite{b_tragaldabaskiel} and\cite{b_tragaldabaslahaya}.

\begin{figure*}
\centering
\begin{minipage}{.49\textwidth}
  \centering
  \includegraphics[width=0.9 \linewidth]{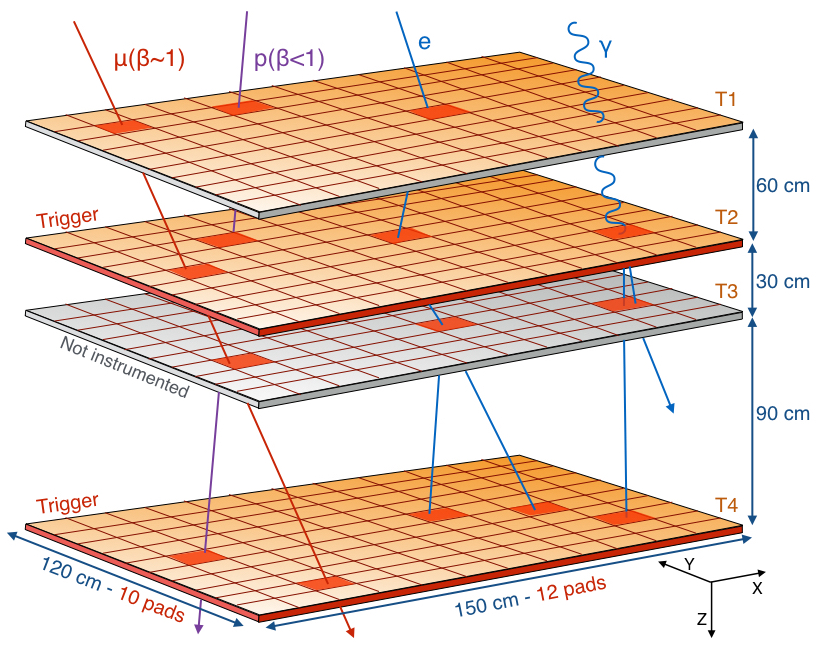}
  \caption{Layout and particle identification capability of TRAGALDABAS. The trigger is provided by planes T2 and T4 in coincidence. The plane T3 is not yet instrumented.}
  \label{f_tragaldabaslayout}
\end{minipage}\hfill%
\begin{minipage}{.49\textwidth}
  \centering
  \includegraphics[width= 0.9 \linewidth]{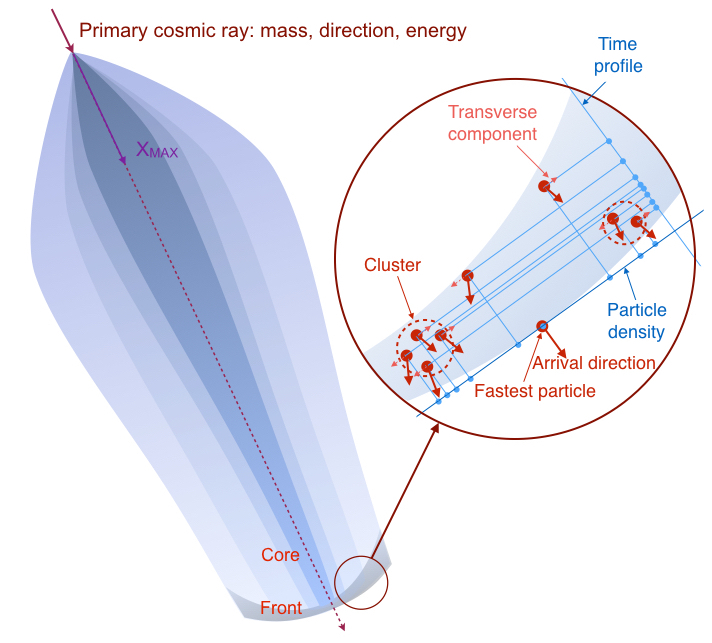}
  \caption{Together with the identification capabilities, TRAGALDABAS will be able to search new observables in the front side of extended air showers.}
  \label{f_easfront}
\end{minipage}\hfill%
\end{figure*}

The TRAGALDABAS detector has four active RPC planes based on the design developed by the LIP of Coimbra for the MARTA extension of the Pierre Auger Observatory\cite{b_marta}. Each plane has an active size of $1.2\times1.5$ m$^2$  and is read out by 120 rectangular pads instrumented with the fast HADES FEE electronics\cite{b_hadesfee}. The data acquisition system is based on the TRBV2 readout board developed at GSI\cite{b_trb}. The acquisition trigger is done by a coincidence between the second and the fourth planes (counting starts from above), hereafter T2 and T4 planes, respectively. The third plane (T3) is not fully instrumented. The detailed layout of the detector and the distance between layers is shown in Figure \ref{f_tragaldabaslayout}. 

The main expected performances of the detector for a charged straight track are:
$\sigma_X \simeq \sigma_Y \sim 3$ cm, $\sigma_T \sim 300$ ps, $\sigma_{\theta} \sim 2.5^{\circ}$, where $\theta$ is the zenith angle. Each plane has two gas gaps of 1 mm width, providing a plane efficiency close to 1. Taking into account the time resolution and the distance between planes we expect a velocity resolution of the order of $\sim 5\%$ of the speed of light. As a consequence, most of the muons and protons can be identified by looking at their velocities (we expect most protons having a velocity below 0.9 c, enough to be separated from muons traveling at $\sim$c). We also expect to be able to make some electron-muon separation during the reconstruction stage by looking
at the quality of a fit to a straight track with a constant velocity. And, as each plane has a material budget of $\sim 0.27 X_0$ (radiation lengths), electrons and gammas could be identified looking at the possible presence of small showers of secondary particles. In order to further improve the muon-electron separation we intend to install, in the near future, a layer of lead of about $3 X_0$ after the third plane, T3.

As shown in Figure \ref{f_easfront}, we also expect to have access to a few other non-standard observables of the EAS as the time and density profiles of the arriving particles, the direction of the fastest particles, the transverse movement components of single particles (or thermalization level of the shower) or the possible presence of clusters of particles, among others. These capabilities might provide new signatures for the identification of cosmic ray air showers.

\section{Data sample}

\begin{figure*}
\centering
\begin{minipage}{.49\textwidth}
  \centering
  \includegraphics[width=0.9 \linewidth]{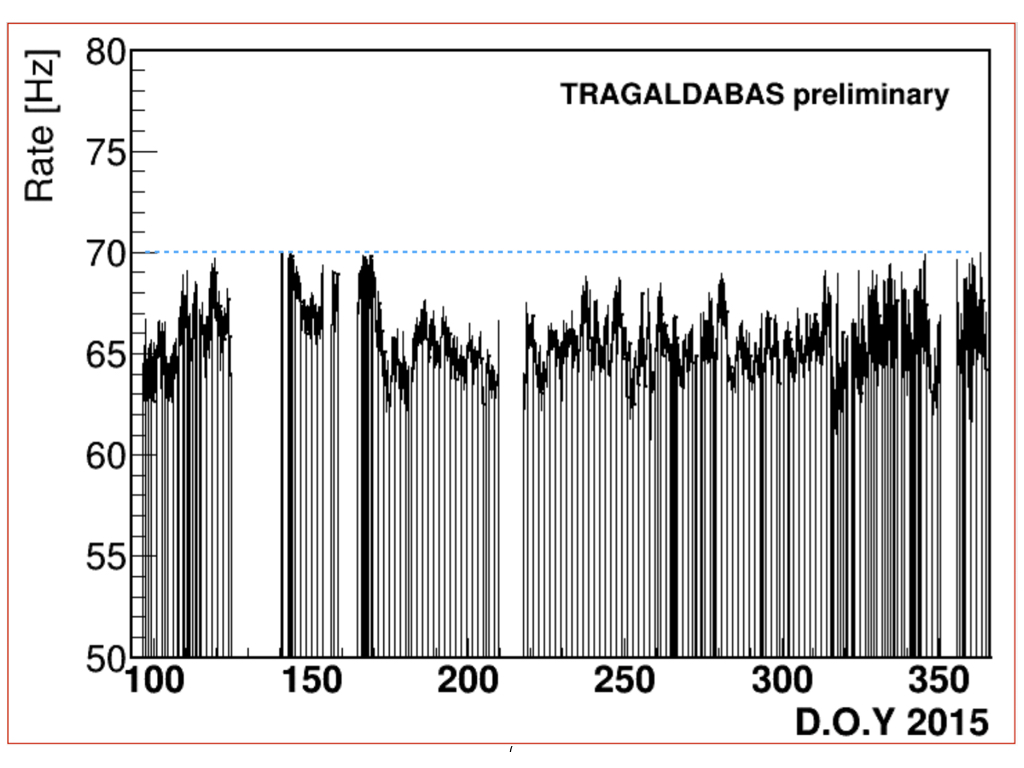}
  \caption{Uncorrected rate of events measured by TRAGALDABAS during 2015.}
  \label{f_tragaldabasrate}
\end{minipage}\hfill%
\begin{minipage}{.49\textwidth}
  \centering
  \includegraphics[width= 0.9 \linewidth]{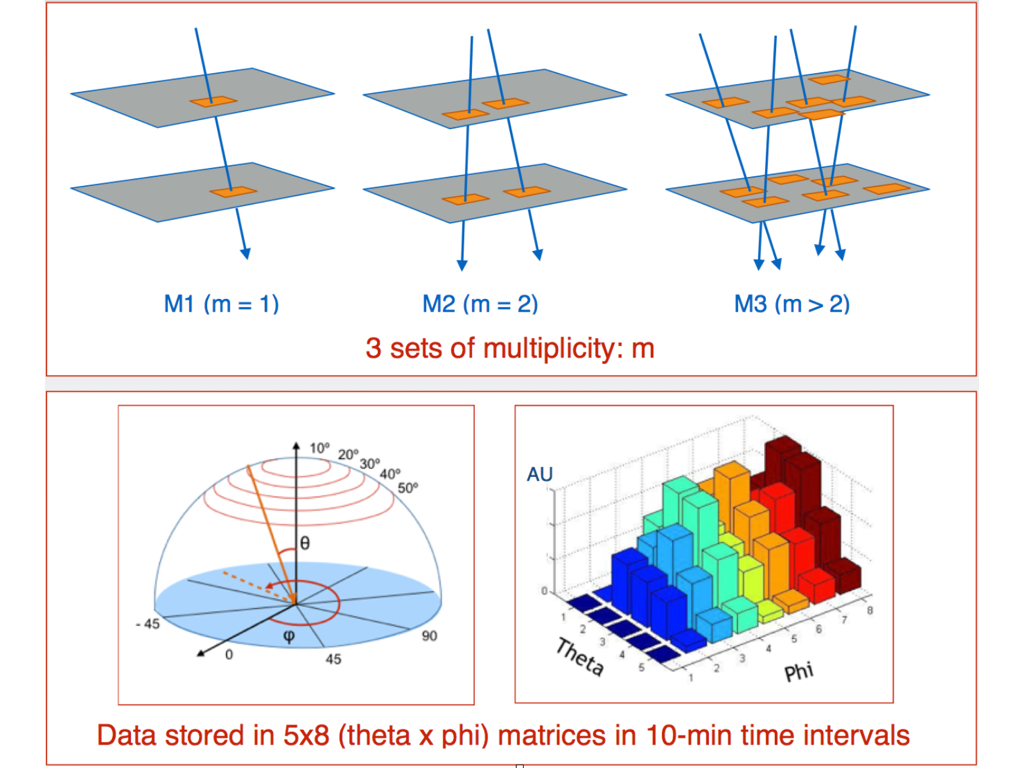}
  \caption{Definition of M1, M2 and M3 events. The events have been lumped together in 10-minute intervals, distributed in 40 $(\theta, \phi)$ cells.}
  \label{f_multiplicitygroups}
\end{minipage}\hfill%
\end{figure*}

The TRAGALDABAS detector has been taking data regularly since April 2015. 
The duty cycle of the detector during that period was about  $90\%$ not including dead periods in the data acquisition either for maintenance or improvements of the detector.
Figure \ref{f_tragaldabasrate} shows the trigger rate of the year 2015. 

Before proceeding with the analysis of the data that will be shown in the next sections, we studied the correlation of the data rate with both the local temperature and barometric pressure. The correlation with the temperature seemed to be almost negligible but there was a very significant anti-correlation with the pressure. This effect was corrected with a logarithmic fit to the first six months period of the collected data. 

For this first analysis we only took into account events, or showers, having the same number of fired cells in both trigger planes and giving rise to unambiguous tracks compatible with having the speed of light. We divided the events as a function of their multiplicity: those having one track, M1, those having two tracks, M2, and those having more than two tracks, M3.  All events from each group  were gathered together in 10 minute intervals and according to their arrival directions. For those events having more than one track, the arrival direction was defined as the direction of the first arriving particle. To classify the arrival direction, the upper solid angle acceptance region was divided in eight $45^{\circ}$ azimuthal angle bins and five $10^{\circ}$ zenith angle bins between $0^{\circ}$ and  $50^{\circ}$. The event selection procedure is summarized in Figure \ref{f_multiplicitygroups}.

\section{Preliminary results}

In order to advance our knowledge of real performances of the detector and to verify its sensitivity to different phenomena that are usually studied with other cosmic ray detectors, we have analyzed some subsamples of corrected data. 

\subsection{Detector performance}

\begin{figure*}[t]
\centering
\includegraphics[width=135mm]{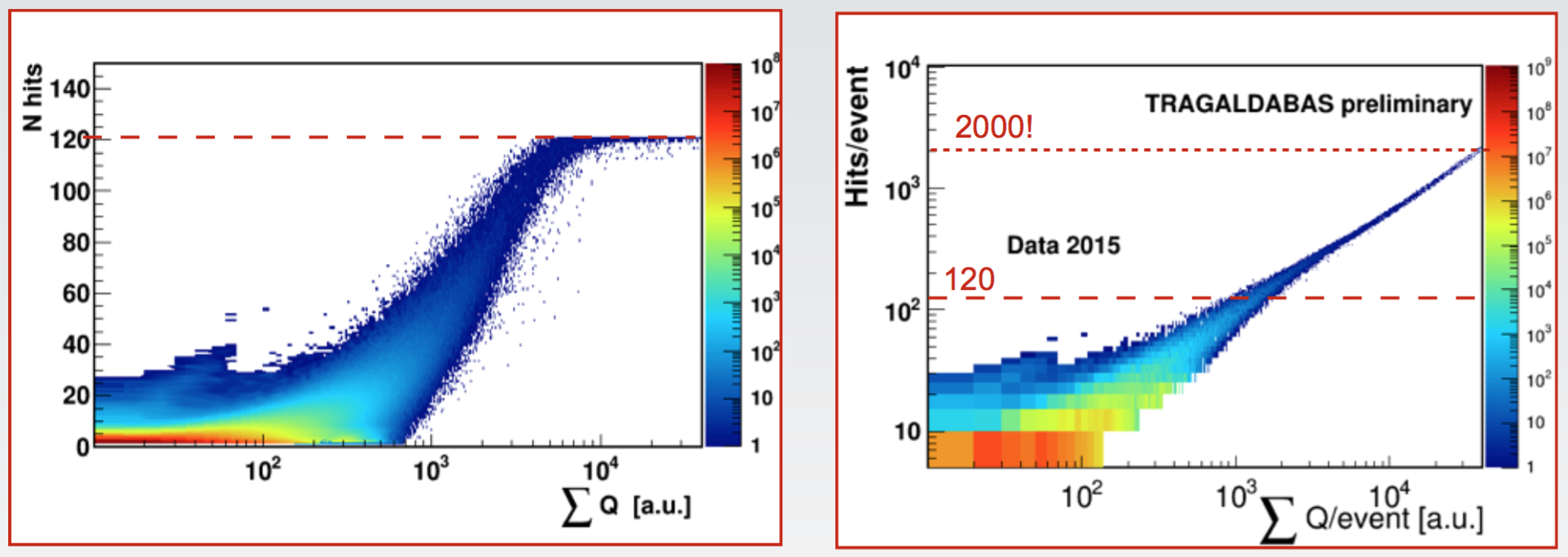}
\caption{
Left) Relationship between the number of fired cells and the total charge measured in the T2 detector.
Right) Estimated number of particles (hits) impinging to the detector assuming that all particles induce the same mean charge on the pad electrodes.
}
\label{f_charges}
\end{figure*}

A very interesting design feature of the detector is that its read-out electronics records the charge induced by the electron-ion avalanches in the electrodes. This means that when two or more particles cross a single cell almost simultaneously, they deposit their charges independently and the total charge collected is almost the sum of the charges induced by each particle independently. As a consequence, although a single plane of the detector has just 120 cells, they have the capability of measuring much higher multiplicities. Figure \ref{f_charges} shows, on the left side,  the number of fired cells versus the total charge collected in the trigger plane T2. On the right side, we show the estimated number of particles per event (axis Y) as a function of the total collected charge on the plane, assuming that all particles deposit the same charge. It may be observed that we might have detected a few showers with up to around two thousands of particles.
The second trend that appears in the lower left part is due to few events with large cross-talk that have not been filtered in the analysis.


\subsection{Solar activity}

During a couple of days in the middle of June 2015 the Sun was very active and produced several consecutive coronal mass ejections, CME, that arrived to the Earth several days later. As a consequence, a very significant Forbush decrease, FD, was observed by almost all the cosmic ray monitors on Earth. The phenomenon was also observed by the TRAGALDABAS detector\cite{b_tragaldabaslahaya}. Making use of the directional capability of our detector, we have carefully analyzed the event trying to better understand  the FD features. With this in mind we have transformed to ecliptic coordinates, centered in the Sun's position, the arrival direction of the cosmic rays reaching the detector. 

\begin{figure*}[t]
\centering
\includegraphics[width=135mm]{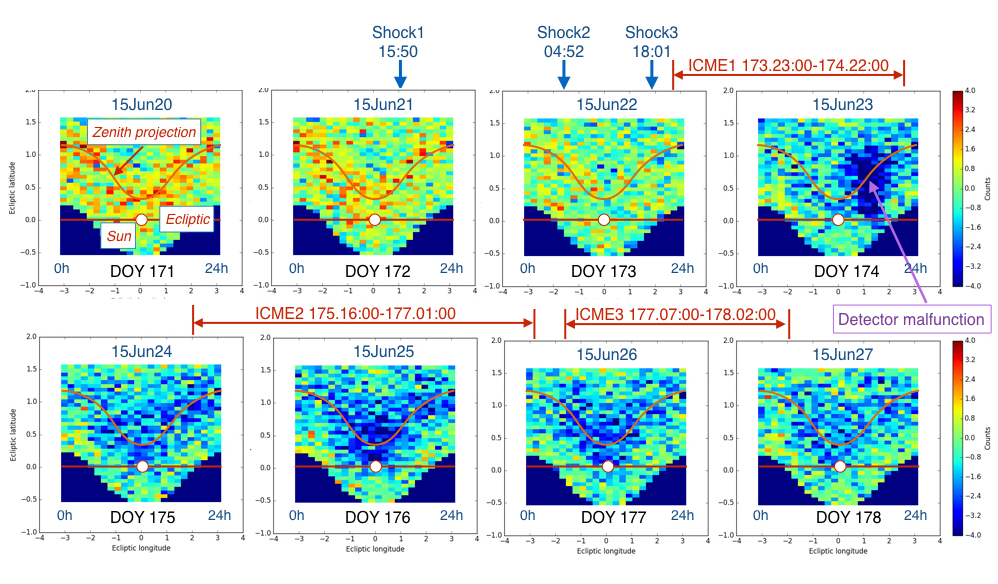}
\caption{Daily pictures of M1 events between June 20th (DOY:171). and June 27th (DOY:178) represented in ecliptic coordinates. The Sun is in the (12h, 0º) position. Each picture represents the difference between the particles arriving that day and a reference mean taken between 12th and 14th of June. A significative decrease of events related to a Forbush decrease starts on June 23rd, reaching its maximum on June 26th. The light green color represents no changes respect to the reference level.}
\label{f_fd}
\end{figure*}

Figure \ref{f_fd} shows the M1 daily pictures of the sky between June 20th. (Day of the year = 171) and June 27th. (DOY = 178). Every picture represents the difference between the distribution of arriving particles per day and a reference distribution calculated as the mean value of the arrival direction of particles between June 12$^{th}$ and June 14$^{th}$, a period of time where the Sun was relatively quiet. 
 The left side of each picture corresponds to 0 hour and the right side corresponds to 24 hour in universal time.The horizontal red line represents the ecliptic plane and the white circle, at 12h, shows the position of the Sun. The small dark spot on top of the Sun corresponds to the position of the North Pole. The orange curve represents the daily trajectory of the vertical of the detector. Both blue dark regions on the bottom correspond to blind acceptance regions. On the right side of each picture a vertical bar shows the color scale of the distribution. Light green corresponds to the zero level. 
The figure also shows the arrival time of three consecutive shocks followed by three ICMEs (Interplanetary coronal mass ejections).

It may be seen that a significant decrease in the arrival of cosmic rays starts on June 23rd shortly after the arrival of the first ICME. Unfortunately, the detector had a short malfunction after noon and a few particles during that period where lost. Concerning the track density, as could be expected, the most significant lack of particles mainly takes place around the position of the Sun at noon. This lack of particles also follows mainly the vertical direction of the detector. The effect starts to slowly disappear after the day 26th. 
\footnote{It is important to emphasize that, as we are representing differences between distribution the geometrical acceptance effect of the detector falls automatically corrected.}. 

\subsection{Geomagnetic field}

\begin{figure*}
\centering
\begin{minipage}{.49\textwidth}
  \centering
  \includegraphics[width=0.9 \linewidth]{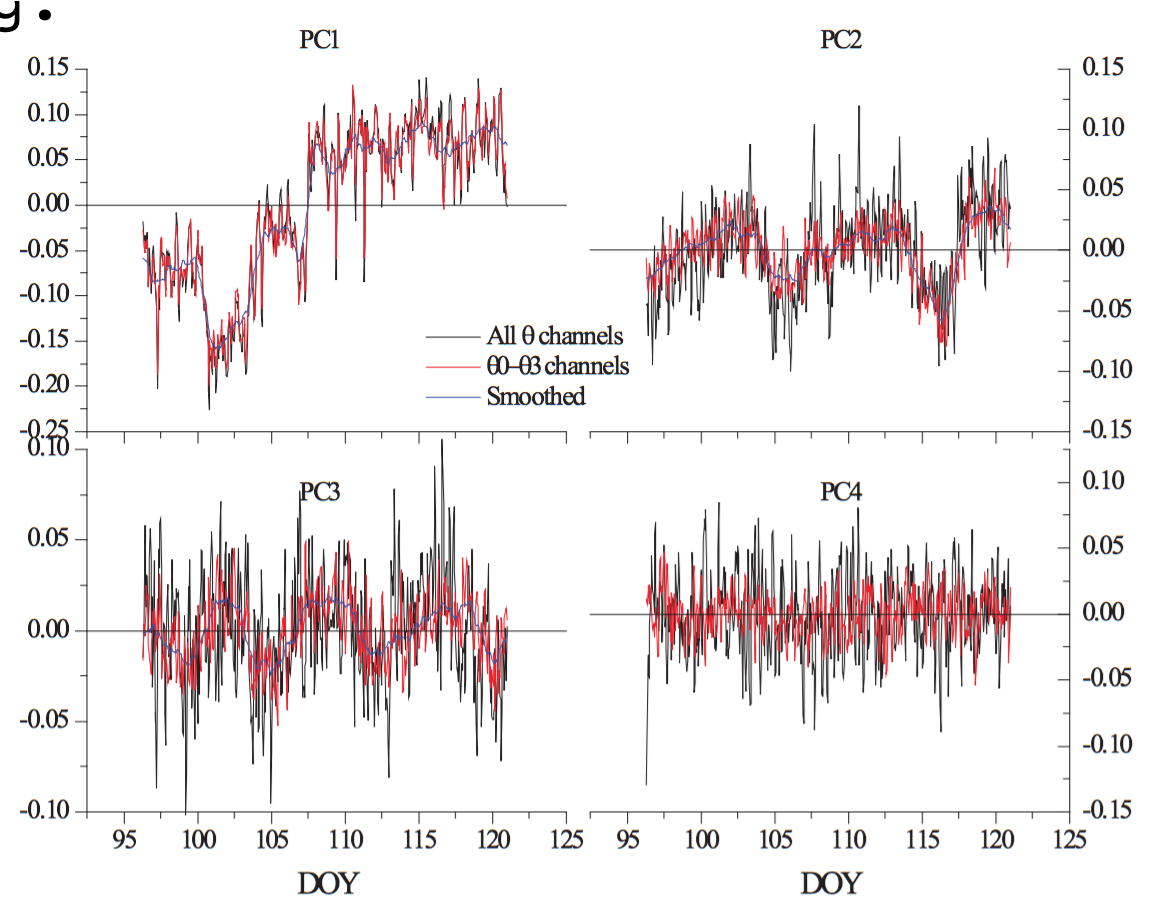}
  \caption{Time behavior of the four principal components of the M1 data. Only the three first components show significative trends.}
  \label{f_pcatrends}
\end{minipage}\hfill%
\begin{minipage}{.49\textwidth}
  \centering
  \includegraphics[width= 0.9 \linewidth]{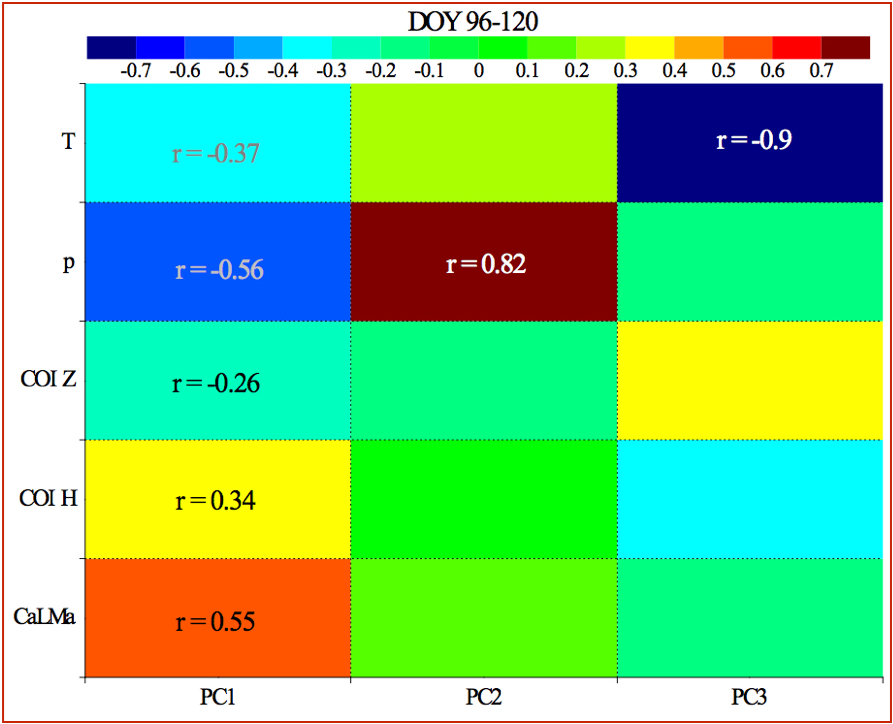}
  \caption{Correlation coefficients  between the three main principal components and several variables: CaLMa neutron rate, Coimbra horizontal (COIZ) and vertical (COIH) magnetic fields, and the pressure (P) and temperature (T) at the laboratory.}
  \label{f_pca}
\end{minipage}\hfill%
\end{figure*}

We have analyzed the possible relationship between the TRAGALDABAS data and both the geomagnetic field (GMF) measured by the CITEUC, located around 300 km south of our detector, and the neutron rate measured by CaLMa neutron monitor, NM, at about 500 km south-east of our detector using a principal component analysis, PCA. Namely, the hourly averaged data for the DOYs (days of the year) from 96 to 120, were submitted to PCA to find common patterns to all channels. Figure \ref{f_pcatrends} show the results for the first four principal components, PC, for M1 data. As can be observed, only the first three components show time variations that differ from white noise. These PCs explain 44-59 $\%$, 5-8 $\%$ and 3-6$\%$ of the data respectively. Figure \ref{f_pca} shows the correlation coefficients between the PCs and the GMF, NM and atmospheric series.  It is interesting to stress that only PC1 shows significative correlations with GMF and NM series. As it explains up to almost 60 $\%$ of  the data, we expect that it could be used for further analysis of the geomagnetic field variations and several space weather events. The remaining two components are mainly correlated with atmospheric variables, pressure and temperature and would be used to disentangle the effects of these variables in the data and, therefore, to improve the estimation of the rates of primary cosmic rays.

\subsection{Atmospheric properties}

The relationship between the cosmic ray measurements and the atmospheric properties, mainly pressure and temperature, have been analyzed almost since the discovery of cosmic rays and the principal intervening phenomena are very well known\cite{b_admitrieva}. Also, many experiments showed the integrated effect of the atmosphere in the measured rate of cosmic rays at Earth's surface. Together with the well known Winter-Summer seasonal effects, very interesting correlations between the rate of very high energy muons and the temperature at the stratosphere have been observed by several experiments such as \cite{b_osprey} and \cite{b_tilav}.

\begin{figure*}
\centering
\begin{minipage}{.44\textwidth}
  \centering
  \includegraphics[width=0.9 \linewidth]{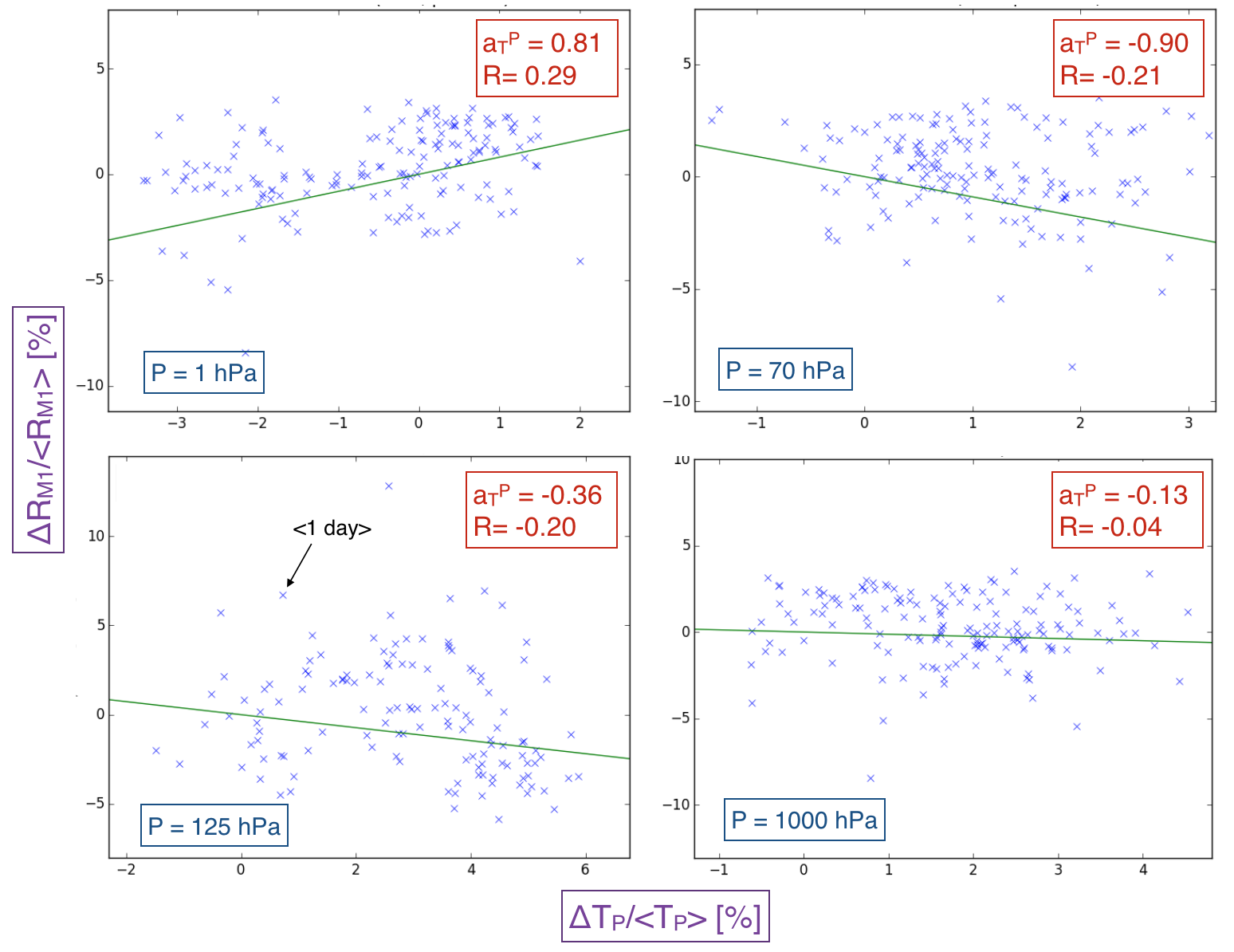}
  \caption{Examples of linear fits of the relative variation of the M1 rate as a function the temperature at different pressure heigths of the atmosphere. }
  \label{f_tslopes}
\end{minipage}\hfill%
\begin{minipage}{.54\textwidth}
  \centering
  \includegraphics[width= 0.9 \linewidth]{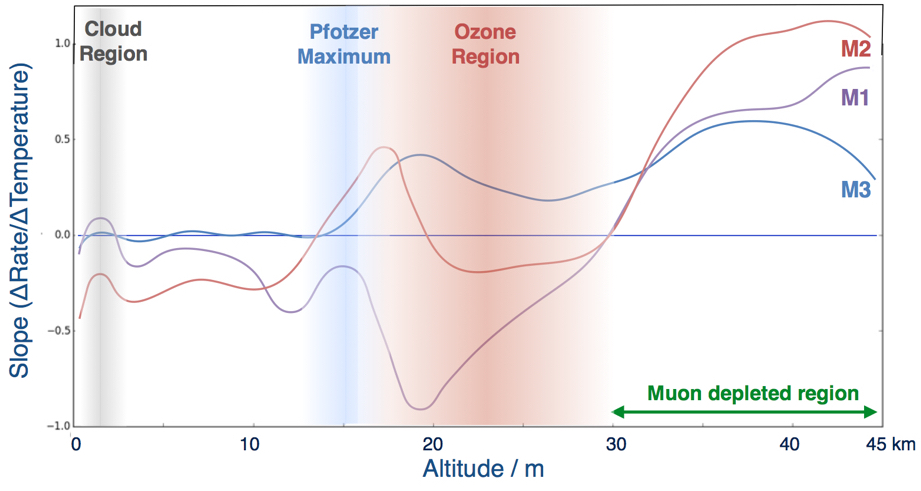}
  \caption{Behavior of the slopes of the linear fits to the relative variation of M1, M2 and M3 rates as a function of the relative variation of the temperature at a given atmospheric height. A few interesting features are commented in the text.}
  \label{f_strends}
\end{minipage}\hfill%
\end{figure*}

In order to determine the effects of the whole atmosphere on our rates, we have analyzed the possible correlations between our measurements and the temperature variations on the different atmospheric layers. To this purpose we have used the reanalysis of the  temperature data provided by ERA-Interim, which belongs to the European Center for Weather Forecasts (ECMWF)\cite{b_era}. Figure \ref{f_tslopes} shows the relative variation of the M1 rate, $\Delta R_{M1}/R_{M1}$, as a function of the relative variation of the temperature, $\Delta T_{P}/T_{P}$ at 45 different pressure levels between P=1hPa (high stratosphere) and P=1000hPa (ground). Linear fits done over the data sets are also shown in the figure. Data corresponds to a six months period, between April and September of 2015, where each point represents the mean value of a single day. Both the value of the slope $\alpha_P$ and the correlation coefficient R among the points are shown in the upper left side of each plot.

We have analyzed if there was any trend in the behavior of the slopes for all the multiplicities. Figure \ref{f_strends} shows the evolution of the slopes of the fits calculated for the three sets of multiplicities in our data after having associated to each pressure level the corresponding height. We can observe a few interesting features. The most significative one is that all slopes have their maximum above around 30 km, the region where primary cosmic rays start to interact with the atmosphere. At lower altitudes all the slopes decrease but all of them have local maxima around the Pfotzer region, where most of the atmospheric showers reach their maximum number of secondary particles. Below the tropopause, at an altitude of around 12-15 km, the slopes stay constant showing, all of them, a small local maximum around an altitude of 2 km, where most of the stratus clouds are originated. 
The positive behavior of the slopes of high multiplicity events in the stratosphere is compatible with the observations reported in references \cite{b_osprey} and \cite{b_tilav}. Certainly, high energy muons measured by those experiments should be produced by high energy primary cosmic rays and an excess of high energy muons should be accompanied by an excess in high multiplicity events, as we are observing.

\section{Conclusions}

We have installed and start up at the Univ. of Santiago de Compostela, in Spain, a high resolution tracking detector of the Trasgo family, the TRAGALDABAS  for the study of cosmic rays. In its present layout, with three active planes, the detector has been taking data since April 2015. An international collaboration has been organized for the regular maintenance and analysis of the collected data. As a many-layer, many-cell tracking detector, the implementation of sophisticated analysis and calibration tools are needed.
Among them, we can mention:
design of an appropriate object oriented software framework, careful analysis of the acceptance and efficiency corrections, development and optimization of tracking algorithms, reconstruction of multiple particle showers, development of particle and shower identification algorithms or a realistic MonteCarlo.
We are on the way. In the meanwhile,
a preliminary analysis has been carried out in order to explore the capability of the detector to be sensitive to the main research areas of the Collaboration, even before reaching the device maximum performance. This preliminary results are very encouraging. 

\bigskip 
\begin{acknowledgments}

Work partially supported by the C. Desarrollo de las Ciencias, of Madrid. 
\end{acknowledgments}

\bigskip 

\end{document}